\documentclass[conference]{IEEEtran}
\IEEEoverridecommandlockouts
\usepackage{cite}
\usepackage{amsmath,amssymb,amsfonts}
\usepackage{graphicx}
\usepackage{textcomp}
\usepackage{xcolor}
\usepackage{subfigure,bm}
\usepackage{algorithmicx}
\usepackage[linesnumbered,ruled,vlined]{algorithm2e}
\usepackage{algpseudocode}
\usepackage{setspace}
\usepackage{multirow}
\usepackage{amsmath}

\usepackage[margin=0.7in]{geometry}

\graphicspath{{figures/}}

\def\BibTeX{{\rm B\kern-.05em{\sc i\kern-.025em b}\kern-.08em
    T\kern-.1667em\lower.7ex\hbox{E}\kern-.125emX}}
\allowdisplaybreaks

\begin{document}
\title{Energy-aware Allocation of Graph Jobs in Vehicular Cloud Computing-enabled Software-defined IoV}


\author{Minghui LiWang$^{1,2}$, Zhibin Gao$^{1,4}$, Seyyedali Hosseinalipour$^{3}$, Huaiyu Dai$^{3}$, Xianbin Wang$^{2}$\\\\
{1 Dept. of Information and Communication Engineering, Xiamen University, Xiamen, China}\\
{2 Dept. of Electrical and Computer Engineering, University of Western Ontario, London, Canada}\\
{3 Dept. of Electrical and Computer Engineering, North Carolina State University, Raleigh, USA}\\
{4 Corresponding author}\\
{Email: \{mliwang@uwo.ca, gaozhibin@xmu.edu.cn, shossei3@ncsu.edu, hdai@ncsu.edu, xianbin.wang@uwo.ca\}}\vspace*{-1em}
\thanks{This work is supported in part by the National Natural Science Foundation of China (grant nos. 61971365, 61871339), Digital Fujian Province Key Laboratory of IoT Communication, Architecture and Safety Technology (grant no. 2010499), the US National Science Foundation (grant nos. ECCS-1444009, CNS-1824518), the Major Research Plan of
the National Natural Science Foundation of China (grant no.
91638204), and the State Key Program of the National Natural
Science Foundation of China (grant no. 61731012).
}}





\maketitle

\begin{abstract}
Software-defined internet of vehicles (SDIoV) has emerged as a promising paradigm to realize flexible and comprehensive resource management, for next generation automobile transportation systems. In this paper, a vehicular cloud computing-based SDIoV framework is studied wherein the joint allocation of transmission power and graph job is formulated as a nonlinear integer programming problem. To effectively address the problem, a structure-preservation-based two-stage allocation scheme is proposed that decouples template searching from power allocation. Specifically, a hierarchical tree-based random subgraph isomorphism mechanism is applied in the first stage by identifying potential mappings (templates) between the components of graph jobs and service providers. A structure-preserving simulated annealing-based power allocation algorithm is adopted in the second stage to achieve the trade-off between the job completion time and energy consumption. Extensive simulations are conducted to verify the performance of the proposed algorithms.
\end{abstract}


\section{Introduction}

Benefiting from the rapid evolutions of the automobile industry, the internet-of-vehicles (IoV) has shown considerable promise in supporting the future connected/autonomous vehicles and intelligent transportation systems (ITS). Innovations related to immersive applications such as autonomous driving and advanced driver assistants provide safety, convenience and entertainments to both drivers and passengers. Furthermore, advances in computing and sensing techonologies facilitate applications with computation-intensive features (e.g., real-time 3D mapping and road sign recognition), which require massive computational resources. Specifically, graph-representation is used to characterize most of the above-mentioned computation-intensive applications: each application \footnote{``job'' and ``graph job'' are hereafter used interchangeably referring to ``application''.} is modeled as a graph, where the vertices (components) represent either data sources or data processing units while the edges describe the dependency (data flows) between the vertices~\cite{1,2}. 

However, individual on-board equipments usually have difficulties handling such applications owing to the constraints in computational resources and capabilities. As a result, collaborative computing among neighboring vehicles is proved to be helpful in IoV to take better utilization of available surrounding resources while meeting the application requirements of users. Vehicular cloud computing (VCC) technology has emerged as an efficient way of collaborative computing and the enhancement of edge computing, where vehicles can form a cloud (vehicular cloud, VC) via vehicle-to-vehicle (V2V) communications, sharing surplus resources with vehicles who face heavy workloads and limited capabilities. Nevertheless, vehicular ad-hoc networks (VANETs) have posed challenges to achieve satisfactory resource management in VCC due to the lack of global information gathering and coordination. Fortunately, given flexible programmability that facilitates decoupling of the data plane and control plane, software-defined network (SDN) technology is introduced to enable global information sharing and coordination, flexible management, greater service capabilities provisioning in the next-generation IoV, which is generally known as SDIoV~\cite{3,4,5}. This paper integrates the VCC technology with SDIoV, and introduces a novel VCC-enabled SDIoV (VCC-SDIoV) framework to orchestrate on-board resources, where the feasible job allocation among servers in relevant VC is achieved effectively. 

Several works have been dedicated to the study of the architecture of SDIoV~\cite{3} and its applications~\cite{4,5}. Furthermore, there are existing studies devoted to graph job allocation which can be roughly divided into three categories according to the dynamism of the network topology: static~\cite{1,6}, semi-static~\cite{7,8,9,10,11}, and dynamic~\cite{2,12}. Considering static topologies of users and servers in a cloud-enabled network environment, the authors of \cite{1} presented a framework for energy-efficient graph job allocation in geo-distributed cloud networks, where solutions were provided for data center networks of varying scales. The authors in \cite{6} presented a randomized job allocation algorithm, which stabilizes a system with job arrivals/departures and achieves a smooth trade-off between the average execution cost and job queue length. In networks where the topology of either users or servers is dynamic, a Lyapunov optimization-based dynamic offloading scheme for directed graph jobs was evaluated in \cite{7}, which met requirements related to energy conservation and job execution time. In \cite{8}, each application was modeled as a weighted relation graph, a fast hybrid multi-site computation offloading algorithm was proposed to determine the optimal and near-optimal solutions achieving the weighted minimization of execution cost, energy consumption and execution time, by considering various application sizes. The allocation of parallel jobs involving several independent tasks was explored in \cite{9}, aiming to jointly minimize energy consumption and job completion time. Different from the above-mentioned works, edge clouds served as users in \cite{10}, where the authors studied a VC-based graph task offloading mechanism to minimize overall response time. Applications were
modeled as directed graphs in~\cite{11}, sequential and concurrent task offloading mechanisms were presented to minimize application completion time.

The graph job allocation problem in dynamic network environments has rarely been investigated in literature, as limitations in opportunistic server-user communications and component interdependency pose substantial challenges to allocation mechanism designs. In our previous work \cite{12}, a randomized graph job allocation mechanism based on hierarchical tree decomposition was proposed, which can efficiently solve the allocation problem in a near-optimal way while achieving the trade-off between job completion time and data exchange costs among service providers (SPs). We also assessed a new multi-task offloading problem under graph-representation in \cite{2} by considering potential inter-component competition due to task concurrency. However, energy consumption is becoming a concern especially for electric smart vehicles owning to the future tendency of green ITS, which was not considered in our previous work. Moreover, joint allocation of graph job and transmission power generally leads to a coupling in the optimization problem, which further complicates the mechanism design~\cite{13}.

In this paper, a novel VCC-SDIoV framework is proposed that allows for dynamic resource sharing among vehicles under control of the SDN controller. The applications of the job owner (JO) and the relevant VCs are modeled as undirected weighted graphs and virtual machine (VM)-based representation is utilized to quantify available resources on SPs. The joint allocation on transmission power and graph job is studied, through which, each component of the job can be efficiently mapped to a feasible VM of a SP in the related VC. Our main contributions are summarized as follows:

\noindent 1. A novel VCC-SDIoV framework is proposed where the SDN controller on edge computing server can decouple the data plane from the control plane and capture global traffic information of the system. Vehicles are able to effectively share unused resources or enjoy computing service under SDN controller's management.

\noindent 2. The formulation of the energy-aware graph job allocation is proposed to achieve a trade-off between the JO’s job completion time and energy consumption while abiding by the structure-preservation constraints in job and VC graphs.

\noindent 3. The problem turns out to be a nonlinear integer programming (NIP) problem, which is  NP-hard. Moreover, the related constraints rely on addressing the subgraph isomorphism problem. Thus, a structure-preservation-based two-stage allocation scheme is proposed to solve this problem efficiently by decoupling the template searching stage from the power allocation stage. For the former, an hierarchical tree-based random subgraph isomorphism mechanism is applied, by which feasible mappings between the job components and SPs' idle VMs are obtained. For the latter, a structure-preserving simulated annealing-based power allocation approach is proposed to approximate the global optimum and achieve the trade-off between the job completion time and energy consumption.

\noindent 4. Through simulations, we demonstrate that the proposed two-stage allocation scheme in VCC-SDIoV can achieve considerable performance while outperforming baseline methods for various job types and VC structures.

\section{Framework of VCC-SDIoV and System Models}

\subsection{The VCC-SDIoV framework}

Benefiting from software-defined and virtualization technologies, the VCC-SDIoV framework shown in Fig.~1 features the
decoupling of the network control from the data transmission, and monitors the status of vehicles via centralized management. 
The SDN controller on edge computing server have a global view of traffic to efficiently manage available resources and job requirements. The allocation mechanism is proposed based on the VCC-SDIoV architecture, where a configurable resource pool is established for each JO to realize effective component allocation. Specifically, each JO sends its graph job requirement to the SDN controller via road side units (RSUs), after which the SDN controller addresses the joint allocation problem relative to power and the graph job according to the current network status (i.e., the topology of the relevant VC and available resources). Based on the applicable solution, the allocation process is automatically executed and manipulated among vehicles in each relevant~VC.

\begin{figure}[h!t]
\centerline{\includegraphics[width=2.7in,height=5cm]{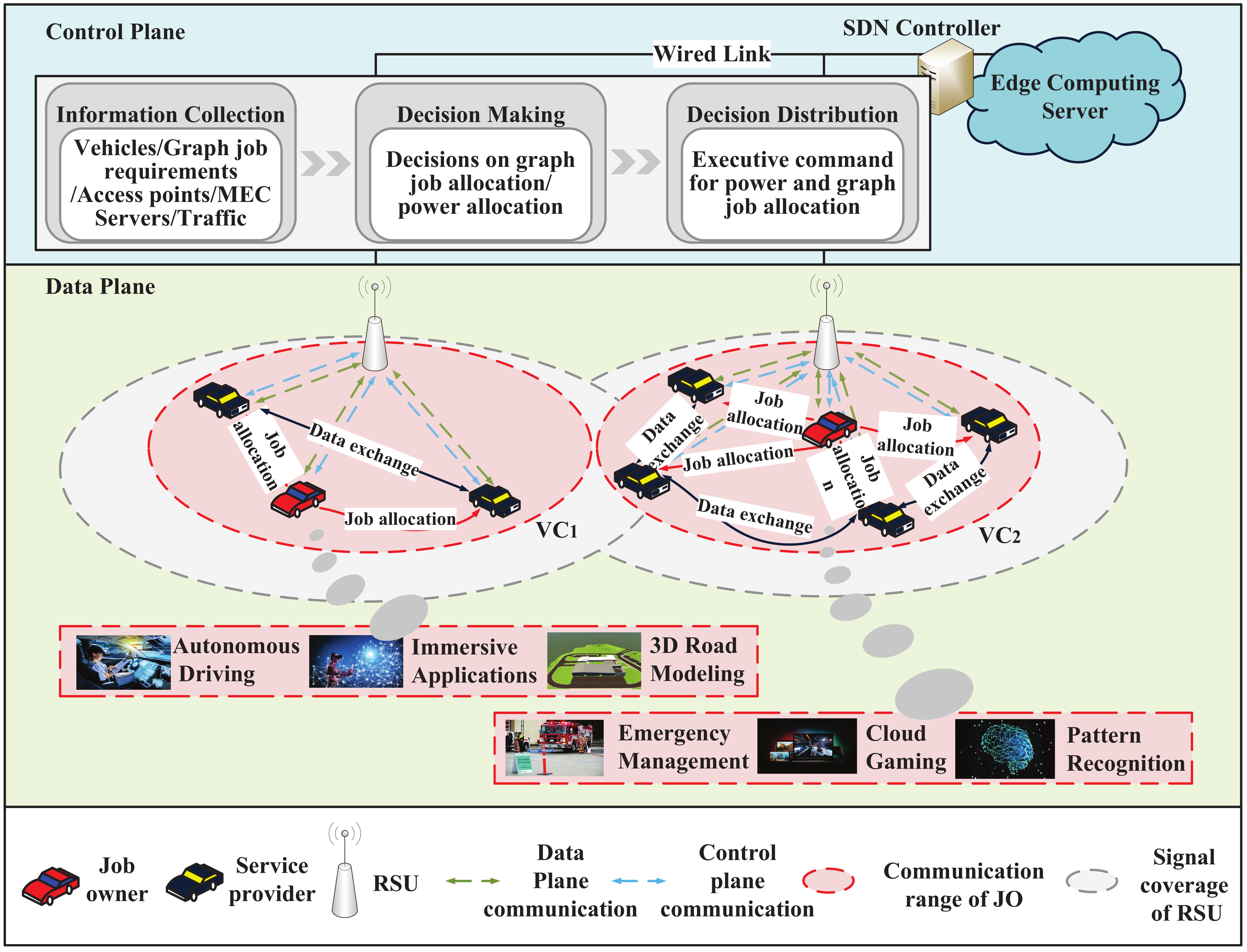}}
\caption{Framework of vehicular cloud computing-enabled software-defined IoV.}
\label{fig1}
\end{figure}

\vspace{-0.7cm}

\begin{figure}[h!t]
\centerline{\includegraphics[width=2.7in,height=3.4cm]{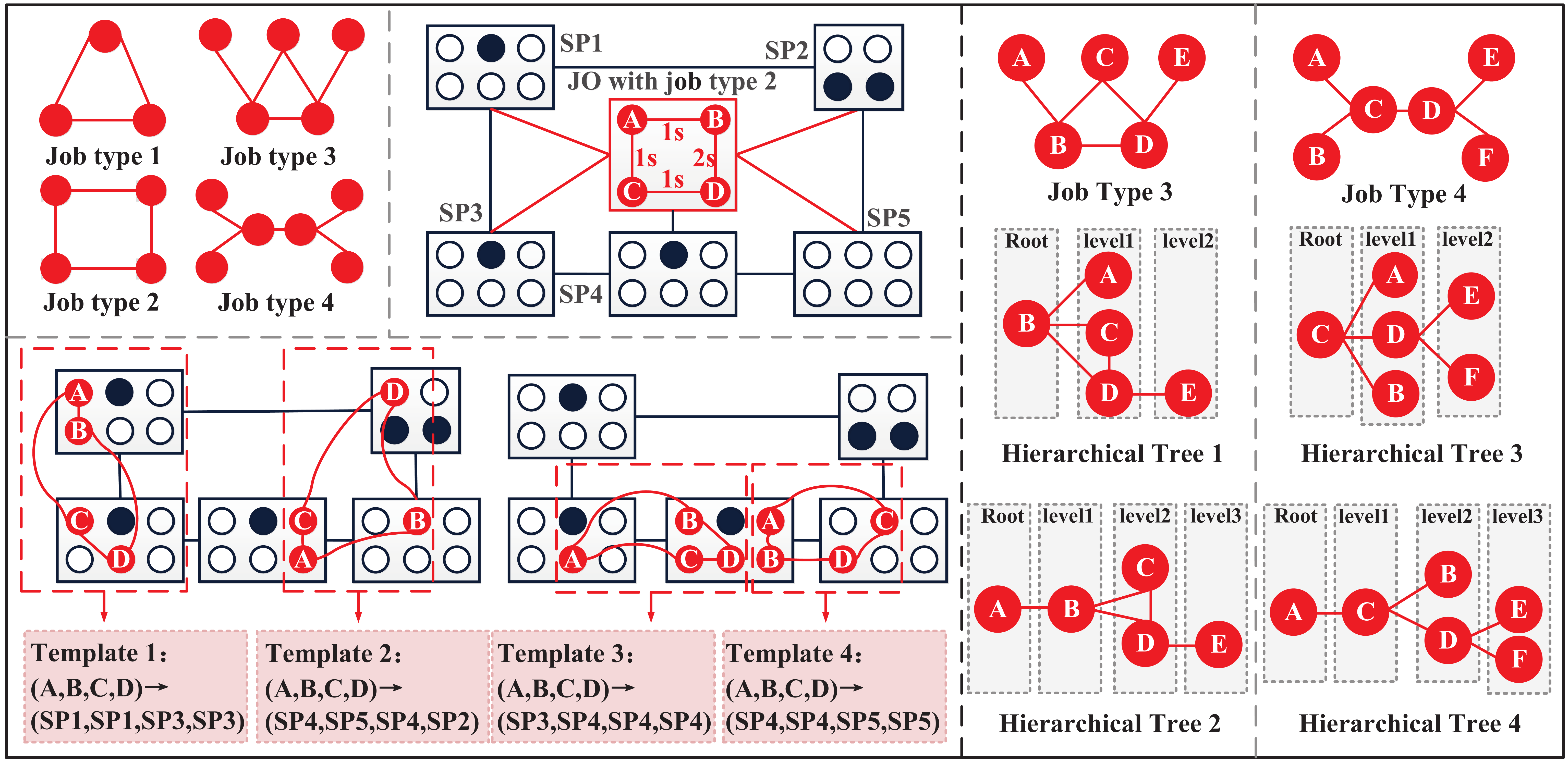}}
\caption{Graph job types and examples of templates and hierarchical trees.}
\label{fig2}
\end{figure}

\vspace{-0.3cm}
\subsection{Vehicular cloud model under graph-representation}

In this paper, we focus on the case where each VC contains one JO and several SPs within JO's communication coverage, where the interference among different VCs is not considered for simplicity. Suppose a VC covers a region containing a JO $o$ and SP set $\bm{S}$, where each SP $s_m\in \bm{S}$ owns different number of fully connected idle VMs for leasing, with the computational capability related to execution time $t^{exec}$ for processing one component. Notably, each available VM can run one component of a graph job at a time. Consequently, a VC is represented as a graph ${\bm{G}}^{\bm{s}}=({\bm{V}}^{\bm{s}}, {\bm{E}}^{\bm{s}}, {\bm{W}}^{\bm{s}})$ containing a set of SPs ${\bm{V}}^{\bm{s}}=\{s_m|s_m\in \bm{S}\}$, each $s_m$ has ${\mathcal{V}_m}$ VMs. The edge set ${\bm{E}}^{\bm{s}}=\{e^s_{mm'}|s_m\in \bm{S},  s_{m'}\in \bm{S}, m\neq m' \}$ denotes the one-hop communication between $s_m$ and $s_{m'}$, while the related weight set ${\bm{W}}^{\bm{s}}=\{{\lambda}_{mm'}|s_m\in \bm{S}, s_{m'}\in \bm{S}, m\neq m'\}$ describes the corresponding parameters of the exponential distribution of contact duration between vehicles which is presented in Section II-D. Fig.~2 shows an example of VC structure.

\subsection{Graph job model and templates}

Consider the job of the JO $o$ in a VC as a graph ${\bm{G}}^{\bm{o}}=({\bm{V}}^{\bm{o}}, {\bm{E}}^{\bm{o}}, {\bm{W}}^{\bm{o}})$ containing a component set ${\bm{V}}^{\bm{o}}=\{v_n|n\in \{1,2,\cdots, |{\bm{V}}^{\bm{o}}|\}\}$, the data size of every $v_n$ is denoted as $d_n$ (bits); and a set of edges ${\bm{E}}^{\bm{o}}=\{e^o_{nn'}|n\in \left\{1,2,\cdots, \left|{\bm{V}}^{\bm{o}}\right|\right\},  n'\in \left\{1,2,\cdots, \left|{\bm{V}}^{\bm{o}}\right|\right\}, n\neq n'\}$ with associated weights ${\bm{W}}^{\bm{o}}=\{{\omega}_{nn'}|n\in \left\{1,2,\cdots, \left|{\bm{V}}^{\bm{o}}\right|\right\},  n'\in \left\{1,2,\cdots, \left|{\bm{V}}^{\bm{o}}\right|\right\},  n\neq n'\}$. The edges represent required data flows between components and the weight ${\omega}_{nn'}$ of edge $e_{nn'}$ indicates the requested connect duration between $v_n$ and $v_{n'}$ for intermediate data interaction. Clearly, the contact duration of SPs that handle these components should be equal to or larger than ${\omega}_{nn'}$. Graph ${\bm{G}}^{\bm{o}}$ represents how the computation split among the components in ${\bm{V}}^{\bm{o}}$ and describe the internal dependency of the components. Various graph job types are presented in Fig.~2.

\smallskip
\noindent
\textbf{Definition 1 (Template):}
For any graph job type, there are several ways (an exponentially large number) in which the job can be mapped over a VC. A template corresponds to one possible mapping in which a graph job can be mapped, examples of which are shown in Fig.~2.


\subsection{Communication model}

The contact duration between vehicles obeys an exponential distribution~\cite{14} with parameter ${\lambda}_{so}$; therefore, the probability of the contact duration $\Delta {\tau}_{so}$ between vehicles $s$ and $o$ exceeding a certain period $T$ is given by $P(\mathit{\Delta}{\tau}_{so}>T|{\lambda}_{so})=e^{-T{\lambda}_{so}}$. The larger the value of $P(\mathit{\Delta}{\tau}_{so}>T| {\lambda}_{so})$, the more assurance can be achieved to protect the required data transmission between moving vehicles. 

As can be construed, each component of a graph job has to be assigned to an idle VM of a SP, while allocation of all the components of the graph job, considering their inherent communication requirements is necessary for successful execution. Let $a_m\in \{0\}\cup \bm{\mathcal{P}}$ denote a fraction of $o$'s transmission power for data delivery to $s_m$, where $\bm{\mathcal{P}}=\left\{k\Delta p\right|k\in \{1,\cdots, \left|\bm{\mathcal{P}}\right|\}\}$ and $\left|\bm{\mathcal{P}}\right|\Delta p=q_o$ in which $\Delta p$ and $q_o$ denotes the the quantization level and a JO's total transmission power, repectively. Specifically, $a_m=0$ indicates that the JO will not use any computing service provided by $s_m$. Given the power allocation profile $\bm{\mathcal{A}}=\{a_m|s_m\in \bm{S}\}$ for SPs in a VC, considering a fading channel between vehicles, the signal to noise ratio (SNR) at $s_m$ at distance $r_{o, s_m}$ away from the JO is given by:

\begin{align}
\label{eq1}
{SNR}_m={Z^2\beta a_m}/{(r^{\alpha}_{o, s_m}\overline{{\mathcal{W}}_0})},
\end{align}
where $Z$ is the fading coefficient, $\beta $ denotes the unitless constant that depends on antenna characteristics and the average channel attenuation, $\alpha $ represents the path loss factor and $\overline{{\mathcal{W}}_0}$ is the background noise power. Here, $\beta ={{\mathcal{G}}_o{\mathcal{G}}_{s_m}w^{2}}/{{(4\pi r_0)}^2}$ where ${\mathcal{G}}_o$ and ${\mathcal{G}}_{s_m}$ are the gain of $o$ and $s_m$ antennas, and $r_0$ and $w$ is the reference distance close to JO and the wavelength, respectively~\cite{15}. Assume $E[Z^2]=1$, the average data transmission rate $r_m$ from $o$ to $s_m$ is defined as (2) where $\mathcal{W}$ denotes the channel bandwidth. The channel condition is assumed to be stable during the periods of job allocation, while it varies independently from one allocation to another~\cite{16}.
%

\begin{align}
\label{eq2}
r_m (\bm{\mathcal{A}})=\mathcal{W}{\log}_2\left(1+\beta a_m / (r^{\alpha}_{o, s_m}\overline{{\mathcal{W}}_0})\right)
\end{align}


\subsection{Computation and energy consumption model}

Let the binary indicator $x_{nm}=1$ denote the mapping of component $v_n$ to SP $s_m$; $x_{nm}=0$, otherwise. For notational simplicity, let $\bm{\mathcal{X}}={[x_{nm}]}_{1\leq n\leq \left|{\bm{V}}^{\bm{o}}\right|,1\leq m\leq \left|\bm{S}\right|}$, which corresponds to a template. The completion time $t^o$ of the graph job can be calculated based on the slowest processed component of the job, which depends on an unknown transmission power allocation solution:

\begin{align}
\label{eq3}
t^o\left(\bm{\mathcal{X}}, \bm{\mathcal{A}}\right)={\max {\left[\frac{x_{nm}\sum^{\left|{\bm{V}}^{\bm{o}}\right|}_{n=1}{d_n}}{r_m\left(\bm{\mathcal{A}}\right)}\right]}_{1\leq m\leq \left|\bm{S}\right|}}+t^{exec},
\end{align}
where $\sum^{\left|{\bm{V}}^{\bm{o}}\right|}_{n=1}{d_n}$ and $\frac{x_{nm}\sum^{\left|{\bm{V}}^{\bm{o}}\right|}_{n=1}{d_n}}{r_m\left(\bm{\mathcal{A}}\right)}$ represents the total amount of data size and the relevant data transmission time of components from $o$ to $s_m$, respectively. Specifically, the JO would incur extra overhead in terms of energy when transmitting component data to SPs via wireless access. Thus, the energy consumption of $o$ is calculated as:

\begin{align}
\label{eq4}
c^o (\bm{\mathcal{X}}, \bm{\mathcal{A}})=\sum^{|\bm{S}|}_{m=1}{\sum^{|{\bm{V}}^{\bm{o}}|}_{n=1}{x_{nm}\left(\frac{a_m d_n}{r_m (\bm{\mathcal{A}})}\right)}}+\ell,
\end{align}
where $\ell $ indicates the tail energy~\cite{17} given that the JO will hold the channel for a while even after data transmission.


\section{Problem Formulation}

For each VC containing a JO $o$ and SP set $\bm{S}=\{s_m|m\in \{1,2,\cdots, |\bm{S}|\}\}$, the energy-aware graph job allocation problem is formulated as $\mathcal{F}$ in (5), aiming to achieve a trade-off between the job completion time and the JO's energy consumption, under opportunistic contact and resource limitations, where $\bm{T}$ denotes the set of all $|\bm{V}^{\bm{o}}| \times |\bm{S}|$ matrices consisting of binary elements, and $\bm{A}$ is the set of all possible power allocations each consisting of $|\bm{S}|$ elements, each of which belonging to the set $\{0\}\cup \bm{\mathcal{P}}$.
\begin{align}
\label{eq5}
& ~~~~~~~\mathcal{F}: {\mathop{\arg\min}_{\bm{\mathcal{X}} \in {\bm{T}}, \bm{\mathcal{A}} \in  \bm{A}} {{\varepsilon}_1t}^o\left(\bm{\mathcal{X}}, \bm{\mathcal{A}}\right)+{{\varepsilon}_2c}^o (\bm{\mathcal{X}}, \bm{\mathcal{A}})}\\
\textrm{s.t.}\ \notag \\
& \text{C1:}\ \sum^{|{\bm{V}}^{\bm{o}}|}_{n=1}{x_{nm}}\leq {\mathcal{V}}_m, \forall s_m \in \bm{S}, \notag \\
& \text{C2:}\ e^{-\left(\left|\frac{d_n}{r_m (\bm{\mathcal{A}})}-\frac{d_{n'}}{r_{m'}\left(\bm{\mathcal{A}}\right)}\right|+{\omega}_{nn'}\right){\lambda}_{mm'}} \geq {\theta}_1, \text{ if }s_m\neq s_{m'}\notag \\
&~~~~\text{ and } x_{nm}\times x_{n'm'}=1, \forall e^o_{nn'}\in E^o~\text{and}~\forall s_m \in \bm{S}, \notag \\
& \text{C3:}\ e^{-{\lambda}_{mo}\times \sum^{|{\bm{V}}^{\bm{o}}|}_{n=1}{\frac{x_{nm} d_n}{r_m (\bm{\mathcal{A}})}}}\geq   {\theta}_2, \forall s_m \in \bm{S}, \notag \\
& \text{C4:}\ a_m\triangleq 0, \text{if}\sum^{\left|{\bm{V}}^{\bm{o}}\right|}_{n=1}{x_{nm}}=0, \forall s_m \in \bm{S},  \notag \\
& \text{C5:}\  \sum^{|\bm{V^o}|}_{n=1}\sum^{|\bm{S}|}_{m=1}{x_{nm}a_m \leq q_o}. \notag
\end{align}

The objective function in $\mathcal{F}$ refers to the weighted sum of the job completion time and energy consumption with ${\varepsilon}_1$ and ${\varepsilon}_2$ representing the non-negative weight coefficients, that can adjust the trade-off between the job completion time and energy consumption. Constraint C1 imposes restrictions on idle resources for each SP. Constraints C2 and C3 are probabilistic constraints where the former ensures that if two connected components $v_n$ and $v_{n'}$ with edge $e^o_{nn'}$ and relevant weight ${\omega}_{nn'}$ are assigned to different SPs $s_m$ and $s_{m'}$, then the probability of the contact duration between $s_m$ and $s_{m'}$ being larger than $\left|\frac{d_n}{r_m (\bm{\mathcal{A}})}-\frac{d_{n'}}{r_{m'}\left(\bm{\mathcal{A}}\right)}\right|+{\omega}_{nn'}$ should be greater than the threshold ${\theta}_1$, where $\left|\frac{d_n}{r_m (\bm{\mathcal{A}})}-\frac{d_{n'}}{r_{m'}\left(\bm{\mathcal{A}}\right)}\right| $ indicates the effective transmission time difference of components. Notably, the order of more than one components being delivered to the same SP is ignored. Similarly, the latter constraint confirms the successful data transmission of components from the JO to the SPs, where vehicles need to maintain contact to preserve the job structure. Constraint C4 indicates that if no component is mapped to $s_m$, the transmission power allocated to the SP is zero; while C5 prevents the case in which the total allocated power may exceed the a JO's upper limit $q_o$.

Here, $\mathcal{F}$ represents an NIP problem with the existence of binary and integer variables $x_{nm}\in \bm{\mathcal{X}}$ and $a_m \in \bm{\mathcal{A}}$, which is NP-hard, where $x_{nm}$ and $a_m$ are coupled with each other and both need to be optimized. Moreover, the constraints related to $\mathcal{F}$ impose solving the subgraph isomorphism problem to obtain adequate templates, which is known to be NP-complete~\cite{1,2,12,18}. In principle, the optimal solution can be obtained via exhaustive search, which is practically infeasible in this case given the prohibitively large size of the feasible set.  Note that determining the templates of mapping components to SPs results in high computational complexity of $O(2^{\left|{\bm{V}}^{\bm{o}}\right|\times \sum^{\left|\bm{S}\right|}_{m=1}{{\mathcal{V}}_m}})$, where $\left|{\bm{V}}^{\bm{o}}\right|$ and $\sum^{\left|\bm{S}\right|}_{m=1}{{\mathcal{V}}_m}$ indicate the number of components in a graph job and available VMs in the related VC, respectively. Then, for each template, the power allocation needs to be determined in $O(|\bm{\bm{\mathcal{P}}}|^{\mathcal{M}})$ complexity, where $\mathcal{M}$ denotes the number of SPs in the relevant template. Consequently, the system can rarely identify the optimal solutions to reconfigure the IoV extemporaneously, as the running time required to solve large and real-life network cases increases sharply with increasing vehicular density (and with the complexity of the VC topology and job structures). This calls for a low-complexity sub-optimal algorithm to efficiently solve $\mathcal{F}$, which is discussed in the next section.

\section{The Structure-Preservation-based Two-Stage Allocation in VCC-SDIoV}

The significance of preserving the structure of the VC and graph job complicates the simultaneous allocation of transmission power among different SPs and job components. In this section, a structure-preservation-based two-stage allocation algorithm is proposed as an efficient method, namely by decoupling the joint transmission power assignment from the component allocation procedure.

\subsection{Stage 1: Template searching via hierarchical tree-based random sub-graph isomorphism}

%

We first focus on a template searching problem ${\mathcal{F}}_1$ as Stage 1, which stands for the key concern of identifying as many templates as possible for graph job allocation in polynomial time, where a hierarchical tree-based random sub-graph isomorphism (TS-HTRS) mechanism is applied. For each VC, the proposed TS-HTRS mechanism for identifying template set $\bm{\mathcal{T}}=\{ \bm{\mathcal{X}_i}|i\in{1,2,...,|\bm{\mathcal{T}}|}\}$ is formulated as ${\mathcal{F}}_1$ in (6), where a template $\bm{\mathcal{X}_i}=\left\{x^i_{nm}|v_n \in {\bm{V}}^{\bm{o}}, 1 \leq m \leq\mathcal{M}_i \right\}$ contains $\mathcal{M}_i$ SPs denoted as set ${\bm{S}}_{\bm{\mathcal{X}_i}}=\left\{s^i_m|s^i_m\in \bm{S}, 1 \leq m \leq \mathcal{M}_i   \right\}$, the binary variable $x^i_{nm}$ indicates the corresponding mapping between $v_n$ and $s^i_m$ in template $\bm{\mathcal{X}_i}$.
\begin{align}
\label{eq6}
&~~~~~~~~~~~~~~~~~~~~~~\mathcal{F}_1:\bm{\mathcal{T}}\\
\textrm{s.t.}\ \notag \\
&\text{C6:} \ \sum^{|{\bm{V}}^{\bm{o}}|}_{n=1}{x^i_{nm}}\leq {\mathcal{V}}^i_m, \forall s^i_m \in \bm{S_{\mathcal{X}_i}}~\text{and}~\forall \bm{\mathcal{X}_i}\in \bm{\mathcal{T}}, \notag \\
&\text{C7:}~\exists e^s_{mm'}\in {\bm{E}}^{\bm{S}}, \text{ if }x^i_{nm}\times x^i_{n'm'}=1, \forall s^i_m \in \bm{S_{\mathcal{X}_i}}, \notag
\end{align}
where ${\mathcal{V}}^i_m$ is the number of available VMs of $s^i_m$. Constraint C6 imposes restrictions on each SP's idle resources, and C7 ensures that each template will remain consistent with the graph job structure. A hierarchical tree (HT) $\bm{H}$ represents the same graph with ${\bm{G}}^{\bm{o}}$ where components are categorized into levels. Specially, a hierarchical tree can be constructed through the following steps: randomly choose a component in ${\bm{V}}^{\bm{o}}$ as a root (${\bm{level}}_{\bm{0}}$), after which components with $l$-hop (s) connection with the root are considered in set ${\bm{level}}_{\bm{l}}$. The hierarchical tree-based approach can solve the sub-graph isomorphism problem effectively by preserving the graph structure during allocation~\cite{12}, which achieves a low computational complextiy $O(|\bm{V}^o|)$ in each iteration. Examples on different hierarchical trees are depicted in Fig.~2, and the key steps of Stage 1 are summarized in \textbf{Algorithm~1}. Lines 2-9 indicate the randomized component allocation based on the relevant hierarchical tree while preserving edges; lines 10-11 retain different templates.

\subsection{Stage 2: The structure-preserving simulated annealing-based power allocation algorithm}

For a given template $\bm{\mathcal{X}_i}\in \bm{\mathcal{T}}$ obtained from Stage 1, the corresponding near-optimal power allocation solution 
$\bm{\mathcal{A}_i}=\{a^i_m|1 \leq m \leq \mathcal{M}_i \}$ can be obtained by solving problem ${\mathcal{F}}_2$, where $a^i_m$ stands for the transmission power allocated to $s^i_m$:
\begin{align}
\label{eq7}
&\mathcal{F}_2: \bm{\mathcal{A}_i}={\mathop{\arg\min}_{\bm{\mathcal{A}}\in  \bm{A}} {{\varepsilon}_1t}^o\left(\bm{\mathcal{X}_i}, \bm{\mathcal{A}}\right)+{{\varepsilon}_2c}^o (\bm{\mathcal{X}_i}, \bm{\mathcal{A}})} \\
\textrm{s.t.}\ \notag \\
& \text{C8:}\ e^{-\left(\left|\frac{d_n}{r_m (\bm{\mathcal{A}})}-\frac{d_{n'}}{r_{m'}\left(\bm{\mathcal{A}}\right)}\right|+{\omega}_{nn'}\right){\lambda}_{mm'}}\geq {\theta}_1, \text{ if }s^i_m\neq s^i_{m'}\notag \\
&~~~~\text{ and }x^i_{nm}\times x^i_{n'm'}=1, \forall e^o_{nn'}\in {\bm{E}}^{\bm{o}}~\text{and}~\forall s^i_m \in \bm{S_{\mathcal{X}_i}}, \notag \\
& \text{C9:}\ e^{-{\lambda}_{mo}\times \sum^{|{\bm{V}}^{\bm{o}}|}_{n=1}{\frac{x^i_{nm}  d_n}{r_m (\bm{\mathcal{A}})}}}\geq {\theta}_2, \forall s^i_m \in \bm{S_{\mathcal{X}_i}}, \notag \\
& \text{C10:}\ \sum^{\mathcal{M}_i}_{m=1}{a_m \leq q_o}.\notag
\end{align}
Similar to (5), constraints C8 and C9 preserve the relevant weights of edges in ${\bm{G}}^{\bm{o}}$. Specially, C8 ensures the data interaction duration between two different SPs $s^i_m$ and $s^i_{m'}$, required by connected components in a job; C9 prevents the case where component data transmission time may surpass the opportunistic contact duration between JO and SP. Constraint C10 guarantees that the upper limit of a JO's transmission power will not be exceeded.

\begin{algorithm}
{\small
\caption{TS-HTRS (Stage 1)}
\SetKwData{Left}{left}\SetKwData{This}{this}\SetKwData{Up}{up}
\SetKwFunction{Union}{Union}\SetKwFunction{FindCompress}{FindCompress}
\SetKwInOut{Input}{Input}\SetKwInOut{Output}{Output}

\Input{${\bm{G}}^{\bm{o}}$, ${\bm{G}}^{\bm{s}}$, the number of iterations ${i}^*$.}

\Output{the feasible templates set $\bm{\mathcal{T}}$.}

\textbf{Initialization:} $\bm{\mathcal{T}}\leftarrow \emptyset $; $\bm{\mathcal{H}}\leftarrow \emptyset $; ${\bm{\mathcal{X}}}_{\bm{1,0}}\leftarrow \emptyset $; create HT set $\bm{\mathcal{H}}\leftarrow \{{\bm{H}}_{\bm{h}}=(\bm{L}, {\bm{E}}^{\bm{o}})|h\in \{1,2,\cdots , |\bm{\mathcal{H}}|\}\}$, where ${\bm{L}={{\bm{level}}_{\bm{0}}\cup \bm{level}}_{\bm{1}}\cup \cdots \cup \bm{level}}_{|\bm{L}|}$;

\For{$h=\bm{1}~to~\left|\bm{\mathcal{H}}\right|, $ ${\bm{H}}_{\bm{h}}\in \bm{\mathcal{H}}$}{

\For{$i=1$ to ${i}^*$}{
map the component in ${\bm{level}}_{\bm{0}}$ uniformly at random to one of the available VMs on a SP, denoted as the set ${\bm{s}}_{{\bm{level}}_{\bm{0}}}$; ${\bm{\mathcal{X}}}_{\bm{h,i}}\leftarrow ({\bm{level}}_{\bm{0}},  \bm{s_{{level}_0}}$);

\For{$l=1$~to~$|\bm{L}|$}{
map all components in ${\bm{level}}_{\bm{l}}$ uniformly at random to available VMs on a set of SP, denoted as the set ${\bm{s}}_{{\bm{level}}_{\bm{l}}}$, while satisfying constraints \textbf{C6-1} and \textbf{C6-2} shown below:

\textbf{C6-1:} edges between components in ${\bm{level}}_{\bm{l}}$;

\textbf{C6-2:} edges between component in ${\bm{level}}_{\bm{l-1}}$ and ${\bm{level}}_{\bm{l}}$;

${\bm{\mathcal{X}}}_{\bm{h,i}}\leftarrow {\bm{\mathcal{X}}}_{\bm{h,i}}\cup ({\bm{level}}_{\bm{l}},  {\bm{s}}_{{\bm{level}}_{\bm{l}}}$);

$\bm{\mathcal{X}_i} \leftarrow \bm{\mathcal{X}}_{\bm{h,i}}$;
}
\If{${\bm{\mathcal{X}}}_{\bm{i}} \neq \bm{\mathcal{X}_{i-1}}$}{$\bm{\mathcal{T}} \leftarrow \bm{\mathcal{T}}\cup {\bm{\mathcal{X}}}_{\bm{i}}$;}
}}

\textbf{end algorithm}
}
\end{algorithm}

\begin{figure*}[t]\centering
\subfigure[]{\includegraphics[width=1.87in,height=2.46cm]{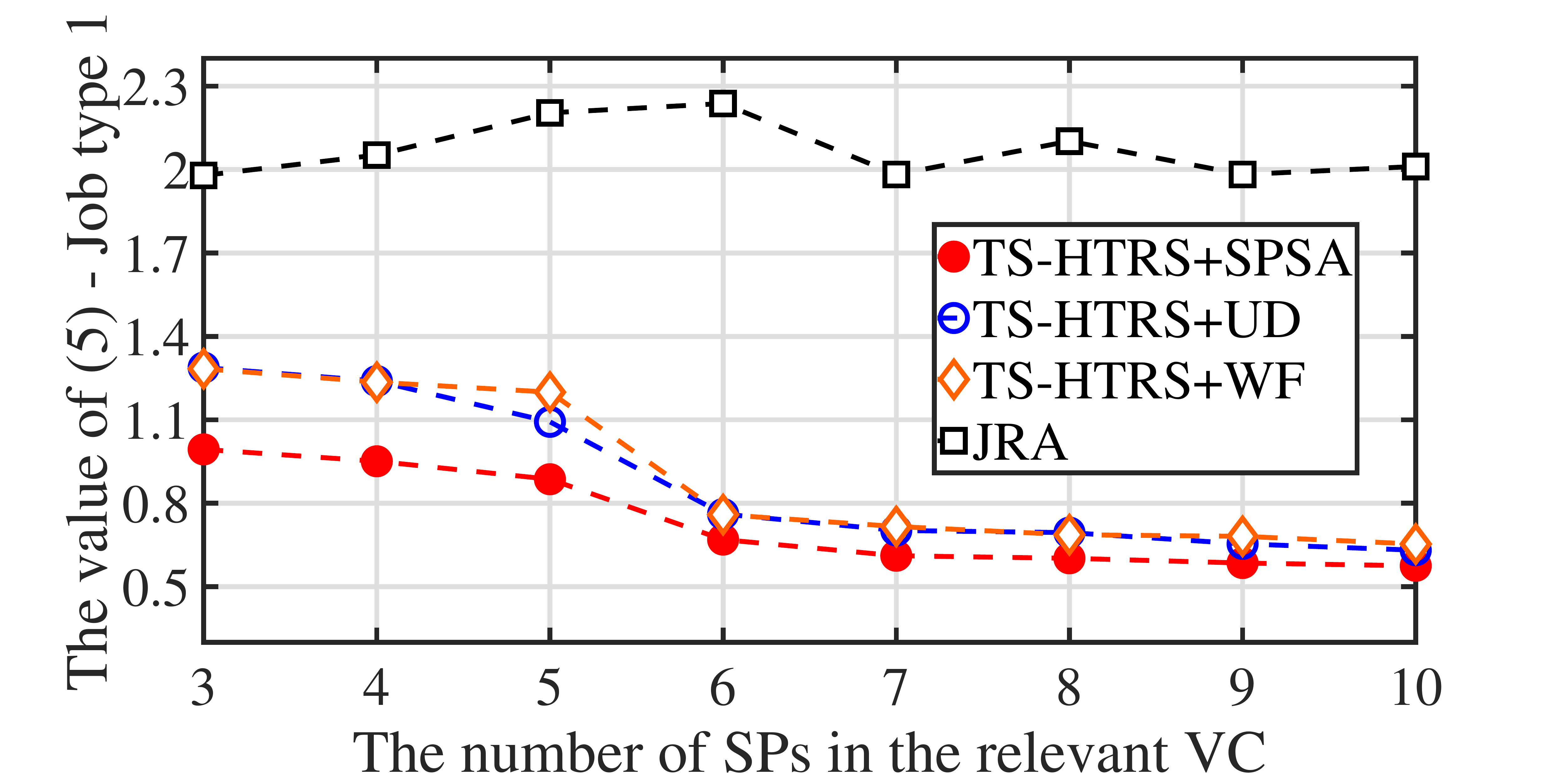}}\hspace{-0.18in}
\subfigure[]{\includegraphics[width=1.87in,height=2.46cm]{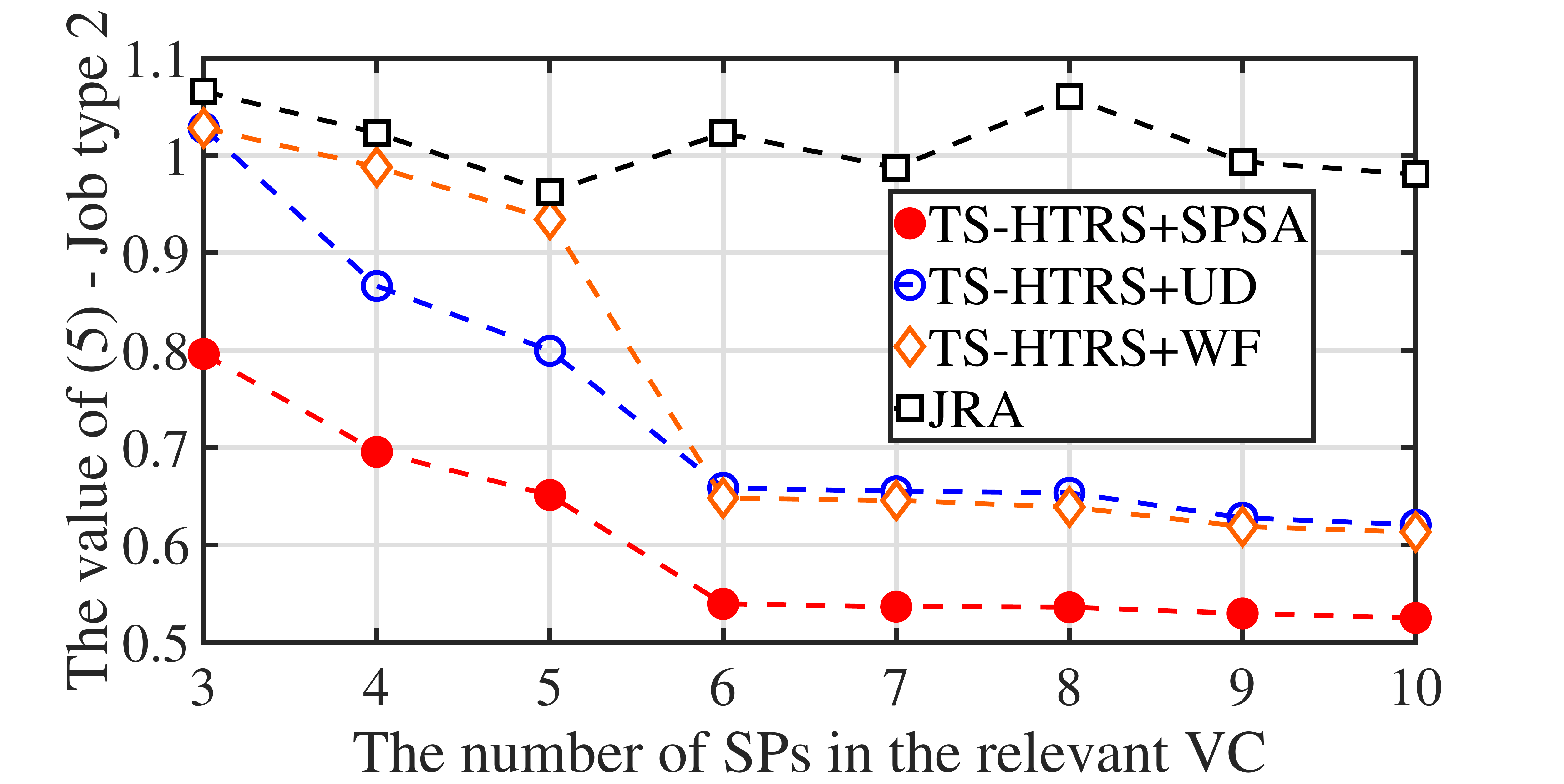}}\hspace{-0.15in}
\subfigure[]{\includegraphics[width=1.87in,height=2.46cm]{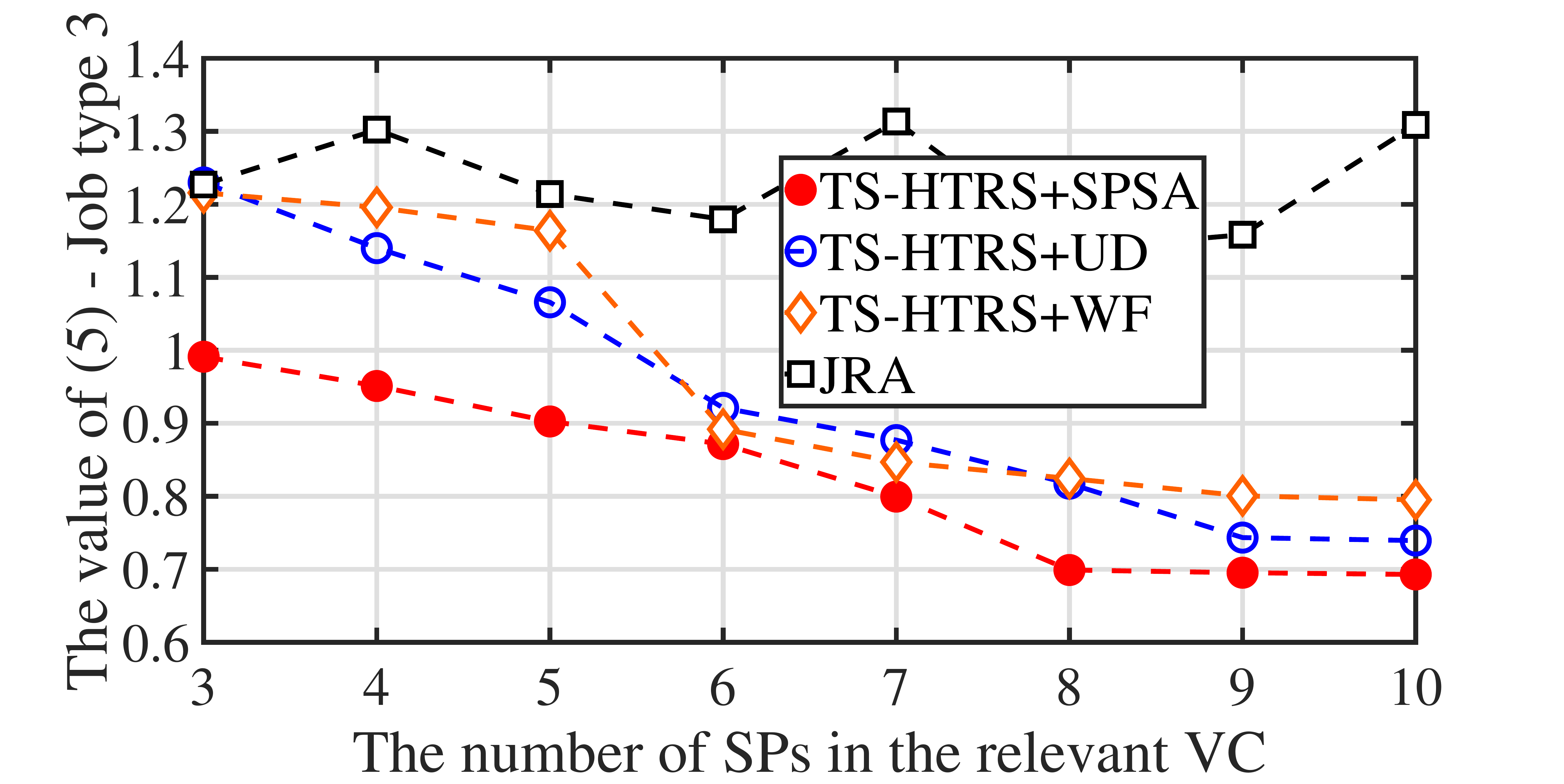}}\hspace{-0.18in}
\subfigure[]{\includegraphics[width=1.87in,height=2.46cm]{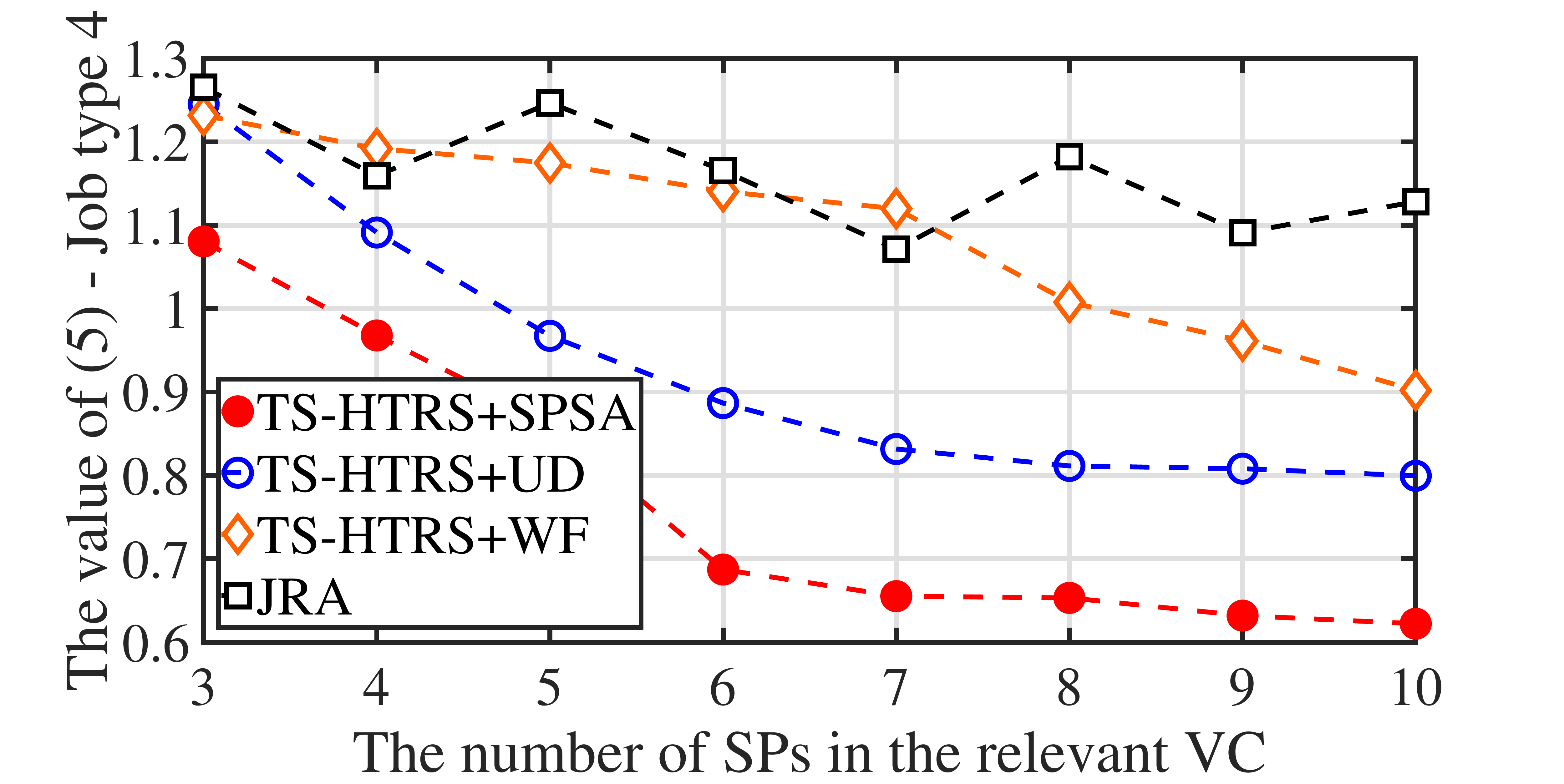}}
\caption{Performance comparisons of the value of objective function (5) for various job types depicted in Fig. 2.}
\label{fig3}
\vspace{-4mm}
\end{figure*}


\begin{figure}[htbp]\centering
\subfigure[]{\includegraphics[width=1.9in,height=2.51cm]{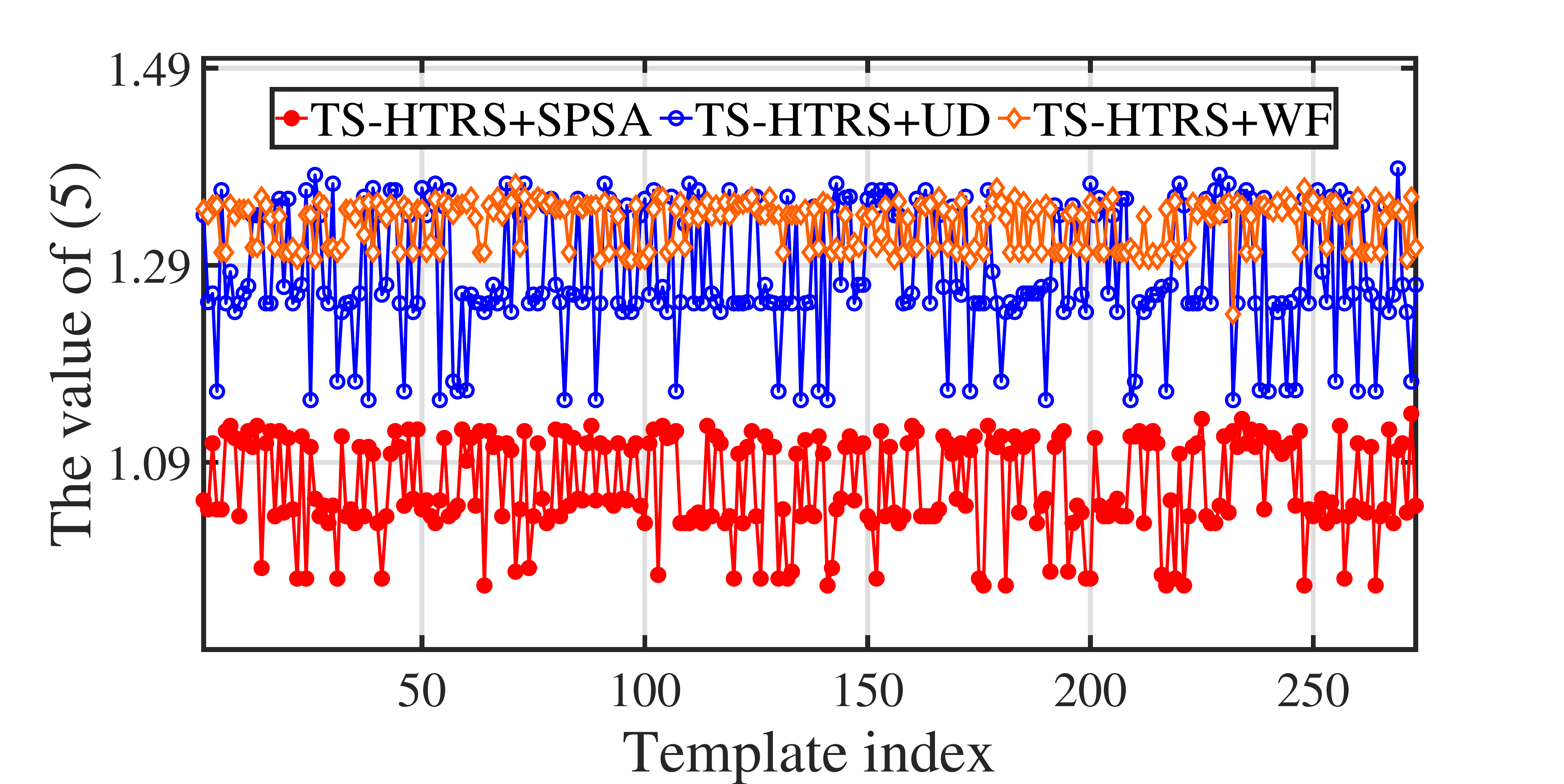}}\hspace{-0.17in}
\subfigure[]{\includegraphics[width=1.7in,height=2.37cm]{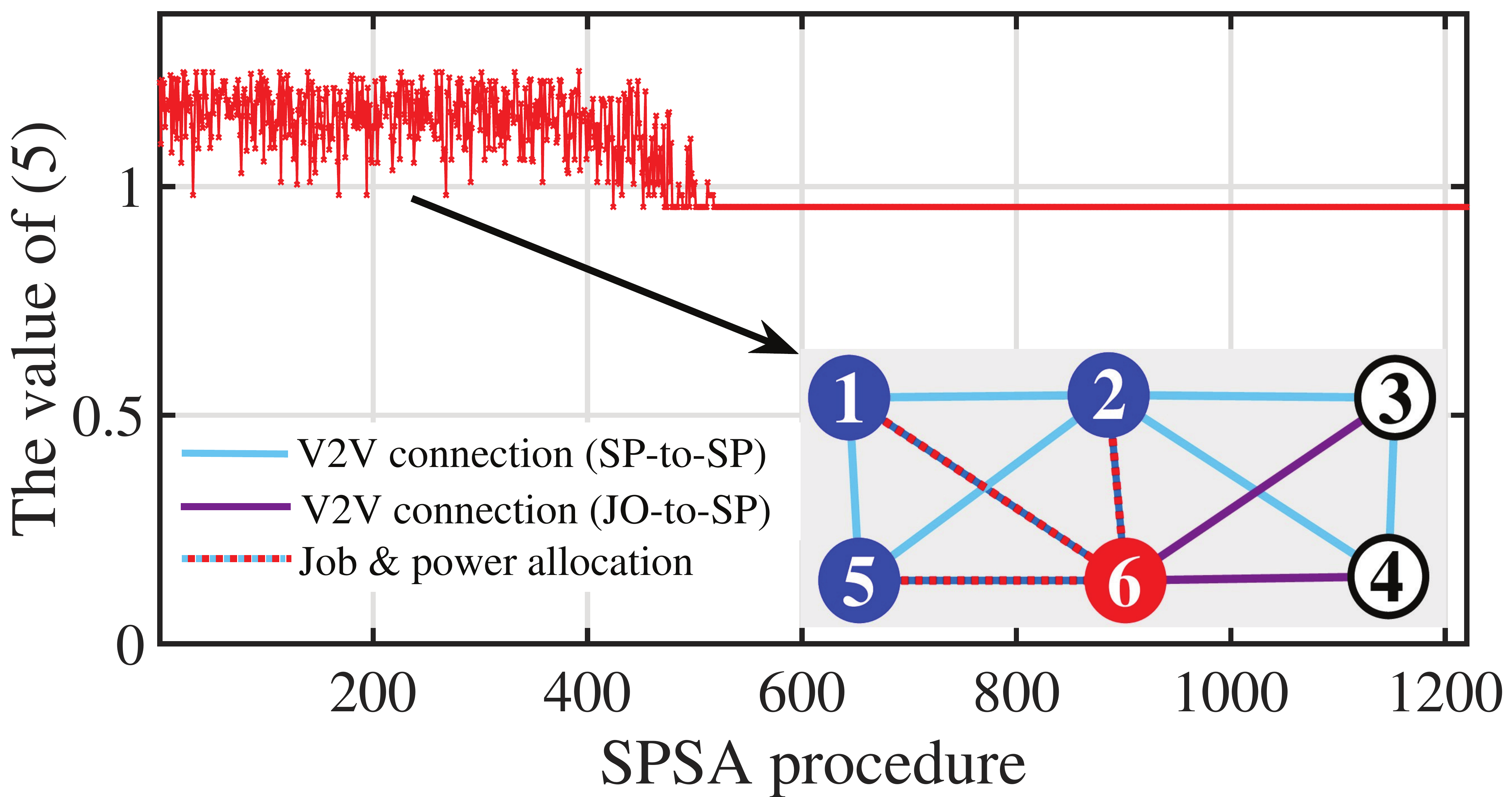}}
\caption{The allocation of job type 1 in a VC containing 5 SPs and 16 available VMs. a) Comparison of the value of (5) for various templates ($x$-axis shows indexes such as the $50^{\text{th}}$ template, the $100^{\text{th}}$ template, etc.); b) An example of the SPSA procedure (number of iterations) for a certain template (1--5: SPs, 6: JO).}
\label{fig3}
\end{figure}

%

${\mathcal{F}}_2$ stands for a NIP problem wherein the power is categorized into various levels as introduced in Section II, which is non-convex. Moreover, the structure-preservation demands in C8 and C9 present obstacles in solving this problem. Therefore, a structure-preserving simulated annealing (SPSA)-based power allocation algorithm~is proposed as an efficient technique of approximating the global near-optimum winthin a large search space, which enables the computational complexity of $O(|\bm{\mathcal{T}}|\times i^{**})$, where $i^{**}$ denotes the number of iterations in SPSA function detailed in \textbf{Algorithm 2}. Based on Markov Chain Monte Carlo and the metropolis criterion, high-quality and cost-effective approximations of near-optimal solutions for this problem can be obtained with a low computational complexity. Given the significance of preserving structures, the approach to create a new state in each iteration of the proposed algorithm is different with the randomized manner in traditional SA. Concretely, for a given template $\bm{\mathcal{X}_i}$ and relevant SP set ${\bm{S}}_{\bm{\mathcal{X}_i}}$, we first calculate the tolerant transmission power (TTP) for the successful delivery of components from JO to each SP, and then adjust the allocation solutions based on TTPs. In other words, the power allocated to $s^i_m$ has to meet the constraint on opportunistic contact duration between JO and $s^i_m$ in C9. Moreover, a new state of power allocation will be abandoned if the contact duration requirement of two connected components cannot be satisfied, given in C8. The pseudo-code of the proposed algorithm is given in \textbf{Algorithm~2} and \textbf{Algorithm~3}, where $\bm{\mathcal{X}_{h,i}}$ denotes the template in the $i^\text{{th}}$ iteration based on the $h^{\text{th}}$ hierarchical tree. For each iteration in \textbf{Algorithm~2}, a new power allocation solution will be created while meeting structure-preservation demands, and compared with the last iteration. Poor solutions can be accepted at a certain probability to skip the local optimum, shown in lines 5-6. The global near-optimal solution is obtained for each different template. In \textbf{Algorithm~3}, the best power and graph job allocation solution can be obtained via comparing the value of objective function among all templates shown in lines 4-8, based on \textbf{Algorithm~2}.

\begin{algorithm}
{\small
\caption{The SPSA function $SrucPresSA$}
\SetKwData{Left}{left}\SetKwData{This}{this}\SetKwData{Up}{up}
\SetKwFunction{Union}{Union}\SetKwFunction{FindCompress}{FindCompress}
\SetKwInOut{Input}{Input}\SetKwInOut{Output}{Output}

\Input{${\bm{G}}^{\bm{o}}$, ${\bm{G}}^{\bm{s}}$, $\bm{\mathcal{P}}$, ${\bm{\mathcal{X}}}_{\bm{i}} \in \bm{\mathcal{T}}$, the number of iterations ${i} ^{**}$, the initial temperature $Temp$.}

\Output{the power allocation solution ${\bm{\mathcal{A}}}_{\bm{i}}$ on template ${\bm{\mathcal{X}}}_{\bm{i}}$.}

\textbf{Initialization:} ${\bm{\mathcal{A}}}_{\bm{i}}\leftarrow \emptyset $; ${\bm{S}}_{\bm{\mathcal{X}_{\bm{i}}}}\leftarrow \left\{s^i_1,s^i_2,\cdots, s^i_{{\mathcal{M}}_i}\right\}$; initial state ${\bm{\mathcal{A}}}_{\bm{i},\bm{1}}$;

\For{$j=1$~to~${i}^{**}$, $Temp>0$}{
Create new state ${\bm{\mathcal{A}}}_{\bm{i},\left(\bm{j+1}\right)}$ based on TTPs while meeting C8, C9 and C10;

$\Delta \tau \leftarrow$~\big(\text{the~value~ of~the~objective~function~of~(5)~given~}
${\bm{\mathcal{A}}}_{\bm{i},\left(\bm{j+1}\right)}~\text{and}~\bm{\mathcal{X}_i}~$\big) $-~$\big(\text{the~value~of~the~objective~function~of~(5)}\\
\text{~given}~${\bm{\mathcal{A}}}_{\bm{i},\left(\bm{j}\right)}$ $\text{and}~\bm{\mathcal{X}_i}$\big);

\If{$\Delta \tau \leq 0$}{${\bm{\mathcal{A}}}_{\bm{i},\left(\bm{j+1}\right)}\leftarrow {\bm{\mathcal{A}}}_{\bm{i,j}}$ at probability $\mu $: $\mu =1-\exp (-|\Delta \tau|/Temp)$;}
$Temp \leftarrow Temp-|\Delta \tau|$;

$j=j+1$;}

${\bm{\mathcal{A}}}_{\bm{i}} \leftarrow {\bm{\mathcal{A}}}_{\bm{i},\bm{j}}$;

\textbf{end algorithm}
}
\end{algorithm}

\vspace{-0.2in}

\begin{algorithm}
{\small
\caption{SPSA-based power allocation (Stage 2)}
\SetKwData{Left}{left}\SetKwData{This}{this}\SetKwData{Up}{up}
\SetKwFunction{Union}{Union}\SetKwFunction{FindCompress}{FindCompress}
\SetKwInOut{Input}{Input}\SetKwInOut{Output}{Output}

\Input{$\bm{\mathcal{T}}$.}

\Output{the near-optimal solution ${\bm{\mathcal{A}}}^*$ and ${\bm{\mathcal{X}}}^{*}$.}

\textbf{Initialization:} ${\bm{\mathcal{X}}}^{*}\leftarrow \emptyset $; ${\bm{\mathcal{A}}}^*\leftarrow \emptyset $; ${value}_0\leftarrow +\infty $;

\For{~$i=1$~to~$|\bm{\mathcal{T}}| $, ${\bm{\mathcal{X}}_{\bm{i}}} \in \bm{\mathcal{T}}$}{${\bm{\mathcal{A}}}_{\bm{i}}\leftarrow SrucPresSA ({\bm{\mathcal{X}}_{\bm{i}}}, \bm{\mathcal{P}}, i^{**}, T)$;

${value}_i\leftarrow\text{value of the objective function $(5)$ given}~{\bm{\mathcal{X}}_{\bm{i}}}\text{ and }{\bm{\mathcal{A}}}_{\bm{i}}$;

\If{${value}_i\geq {value}_{i-1}$}{
${\bm{\mathcal{A}}}^*\leftarrow {\bm{\mathcal{A}}}_{\bm{i}-\bm{1}}$;

${\bm{\mathcal{X}}}^{*}\leftarrow {\bm{\mathcal{X}}_{\bm{i}-\bm{1}}}$;

${value}_i\leftarrow {value}_{i-1}$;}

$i=i+1$;}

\textbf{end algorithm}
}
\end{algorithm}

\vspace{-0.15in}

\section{Numerical results and performance evaluation}

This section presents numerical results illustrating the performance of the proposed structure-preservation-based two-stage allocation algorithm (abbreviated as TS-HTRS+SPSA for notational simplicity), where local computing is not considered. Relevant notations and simulation settings are as follows: $q_o=100 \rm{mWatts}$, $d_n\in [500Kb,600Kb]$, $\alpha =4$; $\overline{{\mathcal{W}}_0}=-130dB$, ${\theta}_1={\theta}_2=0.9$, $t^{exec}\in [0.09s,0.11s]$, $r_{o, s_m}\in [0, 200m]$, ${\mathcal{V}}_m\in \{1,2,3,4\}$, ${\omega}_{nn'}\in [0.2,0.4]$, ${\lambda}_{mm'}\in [0.05,0.06]$, ${\varepsilon}_1=(0,1]/\overline{t^{exec}}$, ${\varepsilon}_2=(0,1]/q_o$, where ${\varepsilon}_1$ and ${\varepsilon} _2$ are normalized due to different units of time and energy consumption and $\overline{t^{exec}}$ is the mean value of $t^{exec}$. Moreover, three baseline methods serve as benchmarks to better evaluate the advantages in this paper, listed below: 

\noindent \textbf{a. TS-HTRS + Uniform distribution (UD):} the proposed TS-HTRS is applied to find templates and the transmission power is uniformly allocated to each SP in the relevant template;

\noindent \textbf{b. TS-HTRS + Water-filling (WF):} the proposed TS-HTRS is applied to find templates and the transmission power is allocated to each SP in the relevant template via WF algorithm;

\noindent \textbf{c. Joint randomized allocation (JRA):} the components and transmission power are randomly allocated simultaneously.

%
%

Performance comparison of the objective function value among baseline methods and the proposed algorithm are presented in Fig.~3 for job types shown in Fig.~2. As the number of SPs in a VC increased, the performance of TS-HTRS+UD, TS-HTRS+WF and the proposed TS-HTRS+SPSA improve due to better allocation options. Moreover, the proposed TS-HTRS+SPSA outperforms other methods owing to the SPSA procedure in each template, whereas JRA always returns a larger value due to the randomized feature.

Comparisons of the value of the objective function for different templates and a SPSA procedure are shown in Fig.~4, taking job type 1 as a representative example. As indicated in Fig.~4 (a), HTRS+SPSA outperforms other methods over 273 templates. Based on the certain template where components are mapped to SP1, SP2 and SP5 respectively, the SPSA procedure is detailed in Fig.~4 (b), where fluctuating values indicate the process of jumping out of local optimal solutions, before converging to the global near-optimum.

\section{Conclusion}

This paper studies the joint allocation problem on power and job under graph representation in VCC-SDIoV. A structure-preservation-based two-stage allocation mechanism is proposed to solve the problem in an effective manner by decoupling the template searching problem and the power allocation problem. The effectiveness of the proposed algorithms is revealed through comprehensive simulations. One potential future research direction could involve jobs modeled by directed weighted graphs.


\end{document}